\documentclass[twocolumn,showpacs,preprintnumbers,amsmath,amssymb]{revtex4}
\usepackage{graphicx}% Include figure files
\usepackage{dcolumn}% Align table columns on decimal point
\usepackage{bm}% bold math
\usepackage[usenames]{color}

\usepackage[normalem]{ulem}

\begin{document}

\preprint{APS/123-QED}

\title{Topological Surface States with Persistent High Spin Polarization across Dirac Point in Bi$_{2}$Te$_{2}$Se and Bi$_{2}$Se$_{2}$Te
}% Force line breaks with \\

\author{K. Miyamoto$^{1}$}

 \email{kmiyamoto@hiroshima-u.ac.jp}

\author{A. Kimura$^{2}$}
\author{T. Okuda$^{1}$}
\author{H. Miyahara$^{2}$}
\author{K. Kuroda$^{2}$}
\author{H. Namatame$^{1}$}
\author{M. Taniguchi$^{1,2}$}
\author{S. V. Eremeev$^{3,4}$}
\author{T. V. Menshchikova$^{4}$}
\author{E. V. Chulkov$^{5,6,7}$}
\author{K. A. Kokh$^{8}$ }
\author{O. E. Tereshchenko$^{9,10}$}

 \affiliation{
$^{1}$Hiroshima Synchrotron Radiation Center, Hiroshima University, 2-313 Kagamiyama, Higashi-Hiroshima 739-0046, Japan
 }

\affiliation{
$^{2}$ Graduate School of Science, Hiroshima University, 1-3-1 Kagamiyama, Higashi-Hiroshima 739-8526, Japan
}

\affiliation{
$^{3}$ Institute of Strength Physics and Materials Science, 634021, Tomsk, Russia
}

\affiliation{
$^{4}$ Tomsk State University, 634050, Tomsk, Russia
}

\affiliation{
$^{5}$ Departamento de F$\acute{\rm i}$sica de Materiales UPV/EHU and Centro de F$\acute{\rm i}$sica de Materiales CFM and Centro Mixto CSIC-UPV/EHU, 20080 San Sebasti$\acute{\rm a}$n/Donostia, Basque Country, Spain
}

\affiliation{
$^{6}$ Donostia International Physics Center (DIPC), 20018 San Sebasti$\acute{\rm a}$n/Donostia, Basque Country, Spain
}

\affiliation{
$^{7}$ Centro de F$\acute{\rm i}$sica de Materiales CFM-Materials Physics Center MPC, Centro Mixto CSIC-UPV/EHU, Edificio Korta, Avenida de Tolosa 72, 20018 San Sebasti$\acute{\rm a}$n, Spain
}

\affiliation{
$^{8}$ V.S. Sobolev Institute of Geology and Mineralogy, Siberian Branch,
Russian Academy of Sciences, Koptyuga pr. 3, Novosibirsk, 630090 Russia
}

\affiliation{
$^{9}$Institute of Semiconductor Physics, Siberian Branch, Russian Academy of Sciences,
pr. Akademika Lavrent$^{\prime}$eva 13, Novosibirsk, 630090 Russia
}

\affiliation{
$^{10}$Novosibirsk State University, ul. Pirogova 2, Novosibirsk, 630090 Russia
}

\date{\today}% It is always \today, today,

\begin{abstract}
Helical spin textures with the marked spin polarizations of topological surface states have been firstly unveiled by the state-of-the-art spin- and angle-resolved photoemission spectroscopy for two promising topological insulators Bi$_2$Te$_2$Se and Bi$_2$Se$_2$Te.
The highly spin-polarized natures are found to be persistent across the Dirac point in both compounds.
This novel finding paves a pathway to extending their utilization of topological surface state for future spintronic applications.
\end{abstract}

\maketitle
%\section{\label{sec:level1}Introduction}
Three-dimensional topological insulators (3D TIs) with massless helical Dirac fermions at the surface in a bulk energy gap induced by a strong spin-orbit coupling have attracted a great attention as key materials to revolutionize current electronic devices~\cite{Fu07, Fu072, Qi08, Hasan10}.
A spin helical texture of topological surface state (TSS), where the electron spin is locked to its momentum, is a manifestation of 3D TI, which is sufficiently distinguished from the {\it real}-spin degenerate Dirac cone in graphene.
This situation promises an effective spin polarized current under an electric field as well as a substantial suppression of backscattering in the presence of non-magnetic impurities.
These new states of matter are also expected to provide fertile ground to realize new phenomena in condensed matter physics, such as a magnetic monopole arising from the topological magnetoelectric effect and Majorana fermions hosted by hybrids with superconductors~\cite{Qi09, Fu08}.

 A number of 3D TIs, such as Bi$_{2}$Te$_{3}$~\cite{Chen_Science_09}, Bi$_2$Se$_3~$\cite{Zhang09, Xia_NatPhys_09, Kuroda101}, and thallium-based compounds~\cite{Yan_EPL_10, Lin_PRL_10, Eremeev_Tl_PRB2011, Sato_PRL_10, Kuroda_PRL_10_TlBiSe2, Chen_PRL_10}, were predicted and experimentally realized.
Among the established 3D TIs, the binary tetradymite compounds, Bi$_{2}$Se$_{3}$ and Bi$_{2}$Te$_{3}$ have been mostly studied because of their relatively large energy gap and the simplest TSS.
The spin-momentum locking feature is experimentally proved at least for the upper-lying TSS by spin- and angle-resolved photoemission spectroscopy with widely spread values of raw spin polarizations (20-80~\%)~\cite{Hsieh09, Xia_NatPhys_09, Souma11, Xu11, Pan11, Jozwiak11}, but clear spin polarizations are obscured near and below the Dirac point ($E_{\rm D}$).
The scanning tunneling spectroscopy for Bi$_{2}$Se$_{3}$ under a perpendicular magnetic field has revealed the Landau levels (LL) with energy distance being proportional to $\sqrt{|n|B}$, where $n$ and $B$ denote the LL index and the magnetic field.
It evidently signifies the existence of the surface Dirac cone.
However, such characteristic LLs are missing below $E_{\rm D}$~\cite{Cheng_PRL_10, Hanaguri_PRB_10}.
These features obviously tell us that topological nature of the material survives only in the upper part of TSS but is no longer available below $E_{\rm D}$ in Bi$_{2}$Se$_{3}$ and Bi$_{2}$Te$_{3}$.
The absence of such a topological nature at TSS below $E_{\rm D}$ restricts its variety of spintoronic applications.
Futhermore, in spite of significant efforts to realize the surface isolated transport, the progress has thus been hampered by too small surface contribution in the total conductance~\cite{Checkelsky09, Butch10, Eto10, Qu10} because the uncontrolled bulk carrier doping takes place in Bi$_{2}$Se$_{3}$ and Bi$_{2}$Te$_{3}$ due to the Se vacancy and the Bi-Te antisite defect.

%% Emergence of ternary tetradymite compounds
Recently, one of the ternary tetradymite compounds, Bi$_2$Te$_2$Se,  where the central Te layer is replaced with the Se layer in Bi$_2$Te$_3$, was shown to be a 3D TI by the ARPES measurement~\cite{Xu10, Arakane12}.
Importantly, the suppression of the bulk conductivity is anticipated because the well-confined Se atoms in the central layer are expected to suppress the Se vacancy as well as the antisite defects between Bi and Te atoms.
Actually, a highly bulk resistive feature in this compound has successfully led to the observation of its surface-derived quantum oscillations in the magnetotransport experiment~\cite{Ren10}. Another ternary compound Bi$_2$Se$_2$Te has also been predicted to be a 3D TI but not experimentally verified yet~\cite{Wang11}.

Here a question arises for these compounds; how are these spin polarized natures maintained in their TSSs?
Realization of the TSS with high spin polarization in the wide energy range across the Dirac point is crucial for the ambipolar gate control of the TI devices~\cite{Wang12, Segawa12}.
One can also manipulate the spin orientations by tuning the electron filling level in the TSS (Fig.1(a)) as expected for the dual gate TI device~\cite{Yazyev10}.
In this Letter, we have unambiguously clarified for the first time by the state-of-the-art spin- and angle-resolved photoemission spectroscopy (SARPES) that ternary tetradymite compounds Bi$_2$Te$_2$Se and Bi$_2$Se$_2$Te hold spin polarized TSSs with the marked spin polarizations.
Importantly, their spin polarized natures in the TSS are found to be persistent even below $E_{\rm D}$.

%\section{\label{sec:level2}Experimental}
 %%%%%%%%%%%%%%%%%%%%%%%%%%%%%%%%%%%%%
\begin{figure}
\includegraphics{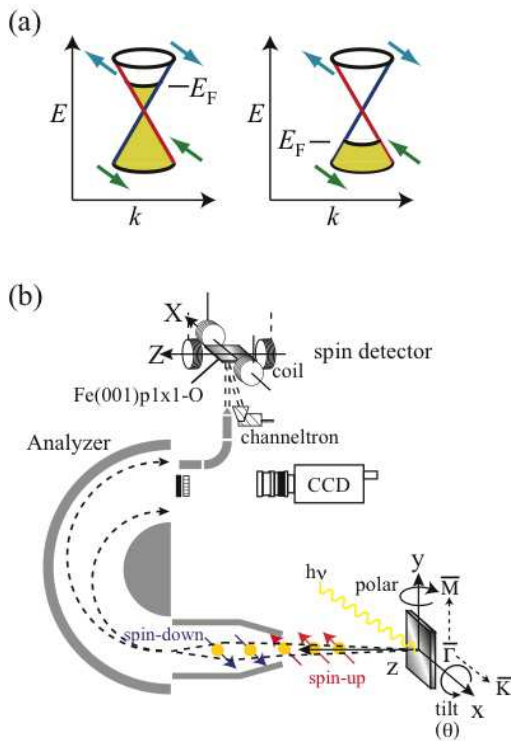}% Here is how to import EPS art
\caption{\label{fig:epsart}(color online) (a) Schematic figure of topological surface state with different filling levels. (b) Schematic figure of our efficient spin resolved spectroscopy (ESPRESSO) machine in HiSOR and the experimental geometry. %(b) Spin-integrated band dispersion of Bi$_{2}$Te$_{2}$Se along $\Gamma_{\rm 1st}$-M- $\Gamma_{\rm 2nd}$ excited with the He discharge lamp ($h\nu$=21.22eV).
}
\end{figure}
%%%%%%%%%%%%%%%%%%%%%%%%%%%%%%%%%%%%%

The crystals of  Bi$_2$Te$_2$Se and Bi$_2$Se$_2$Te were grown from presynthesized mixture of Bi$_2$Te$_3$ and Bi$_2$Se$_3$ by modified vertical Bridgman method \cite{Kokh05}. The purity of elementary Bi, Te and Se for the synthesis of binary compounds was 99.999~\%. The chemical analysis was done by energy dispersive X-ray spectroscopy (EDX).
ARPES  and SARPES experiments were performed with He discharge lamp and synchrotron radiation at Efficient SPin Resolved SpectroScOpy  (ESPRESSO) end station attached to the APPLE-II type variable polarization undulator beamline  (BL-9B) of Hiroshima Synchrotron Radiation Center (HSRC).
The VLEED-type spin polarimeter utilized in the ESPRESSO machine is based on the magnetic target of Fe(001)-p(1$\times$1)-O film grown on MgO(001) substrate, which achieves 100 times higher efficiency compared to those of conventional Mott-type spin detectors~\cite{Okuda11}.
Photoelectron spin polarizations are measured by switching the direction of in-plane target magnetizations.
It simultaneously eliminate the instrumental asymmetry, which is a great advantage for the quantitative spin analysis of non-magnetic systems like in the present case.
This machine can resolve both out-of-plane (Z) and in-plane (X) spin polarization components with high angular- and energy-resolutions as schematically shown in Fig. 1 (b).
  The angle of light incidence was $50^{\circ}$ relative to the lens axis of the electron analyzer.
  The sign of polar (tilt)  angle is defined as positive, in the case of clockwise (anticlockwise) rotation about z-axis (x-axis) as shown in Fig. 1 (b).
  The overall experimental energy- and wavenumber-resolutions of ARPES (SARPES) were set to 25 meV and $<$ 0.008$\rm  \AA^{-1}$ ( 25 meV and $<$ 0.04 $\rm \AA^{-1}$), respectively.
All measurements were performed at a sample temperature of 70 K.
 The samples were $in$-$situ$ cleaved under an ultrahigh vacuum below $1 \times 10^{-8}$ Pa.

DFT calculations were performed using the VASP code
\cite{VASP}, where interaction between the ion cores and valence
electrons was described by the projector augmented-wave method
\cite{PAW}. The generalized gradient approximation was used to
describe the exchange correlation energy. The Hamiltonian contained
the scalar-relativistic corrections, and the spin-orbit coupling was
taken into account by the second variation method. 

 %\section{\label{sec:level3}Results and Discussion}

Figures 2(a) and 2(c) show the ARPES energy dispersion curves of the TSS along $\overline{\rm K}$-$\overline{\Gamma}_{\rm 2nd}$-$\overline{\rm K}$ direction for Bi$_{2}$Te$_{2}$Se and Bi$_{2}$Se$_{2}$Te taken with He discharge lamp ($h\nu$ = 21.22 eV).
We find that the TSSs for both compounds are pronounced rather at $\overline{\Gamma}_{\rm 2nd}$ than at $\overline{\Gamma}_{\rm 1st}$ ({\it not shown}).
 Here,  $\overline{\Gamma}_{\rm 2nd}$ ($\overline{\Gamma}_{\rm 1st}$) denotes the $\overline{\Gamma}$ point in the second (first) surface Brillouin zone (SBZ).
Figures 2(b) and 2(d) highlight intensity maxima of one TSS
branch obtained from momentum distribution curves in the limited
momentum space. One can see that the $E_{\rm D}$ is located at the
binding energy, $E_{\rm B}$ of 415 meV in Bi$_{2}$Te$_{2}$Se and 425
meV in Bi$_{2}$Se$_{2}$Te. The observed dispersions at $E_{\rm
B}$ $<$ $E_{\rm D}$ are almost perfectly linear as demonstrated in
Figs.~2(b) and 2(d). On the other hand, the dispersions at
$E_{\rm B}$ $>$ $E_{\rm D}$ are less steep. The band dispersions for both Bi$_{2}$Te$_{2}$Se and Bi$_{2}$Se$_{2}$Te
are in excellent agreement with that produced by the first principles calculation as shown in Fig.2 (b) and 2 (d).
%%%%%%%%%%%%%%%%%%%%%%%%%%%%%%%%%%%%%
\begin{figure}
\includegraphics{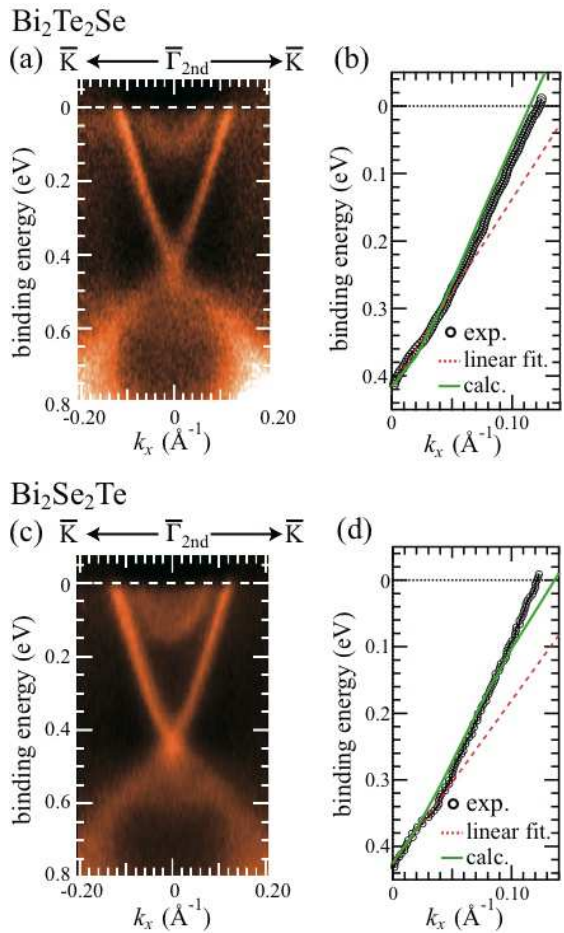}% Here is how to import EPS art
\caption{\label{fig:epsart}(color online) (a), (c) ARPES energy dispersion curves of Bi$_{2}$Te$_{2}$Se and Bi$_{2}$Se$_{2}$Te in $\overline{\rm K}$-$\overline{\Gamma}_{\rm 2nd}$-$\overline{\rm K}$ line. (b), (d) Intensity maxima plot for one TSS branch obtained from momentum distribution curves (open circles) in (a) and (c), respectively. Fitted linear functions are denoted with dashed lines. Energy dispersions of Bi$_{2}$Te$_{2}$Se and Bi$_{2}$Se$_{2}$Te are compared with the results of first principles calculation in (b) and (d). }
\end{figure}
%%%%%%%%%%%%%%%%%%%%%%%%%%%%%%%%%%%%%
\begin{table}
\caption{\label{tab:example}Binding energy at Dirac point ($E_{\rm D}$), group velocity at $E_{\rm D}$ and at Fermi energy ($E_{\rm F}$) of Bi$_{2}$Te$_{2}$Se and Bi$_{2}$Se$_{2}$Te.
}
\begin{ruledtabular}
\begin{tabular}{ccc}
$E_{\rm D}$~[meV]&$v_g~(E_{\rm D})~\times 10^5$~(m/s)& $v_g~(E_{\rm F}~\times 10^5$~(m/s))\\ \hline \\[-5pt]
415$\pm$3&4.2$\pm$0.4&5.8$\pm$0.2\\ \\[-5pt]
425$\pm$3&3.7$\pm$0.2&6.1$\pm$0.2\\

\end{tabular}
\end{ruledtabular}
\end{table}
%%%%%%%%%%%%%%%%%%%%%%%%%%%%%%%%%%%%%

 The group velocity at the Fermi energy (Fermi velocity) is estimated with using the formula 1/$\hbar (\partial$E/$\partial$k) to be ($5.8 \pm 0.2)\times10^{5}~m/s$ in Bi$_{2}$Te$_{2}$Se and ($6.1 \pm 0.2) \times10^{5}~m/s$ in Bi$_{2}$Se$_{2}$Te.
 Also, the group velocity ($v_{g}$) near $E_{\rm D}$ is larger for Bi$_{2}$Te$_{2}$Se than for Bi$_{2}$Se$_{2}$Te as listed in Table I.
The value at $E_{\rm D}$ for Bi$_{2}$Te$_{2}$Se is close to $v_g$=4.6$\times10^5~m/s$ above 130~meV from $E_{\rm D}$ estimated by the magnetotransport measurement~\cite{Ren10}.
Here, the $v_g$ near $E_{\rm D}$ is larger than that for Bi$_{2}$Se$_{3}$ ($v_g$=$2.9\times10^{5}~m/s$ ), while the Fermi velocity is smaller ($v_g$=$6.6 \times 10^{5}~m/s$)~\cite{Kuroda101, Kuroda_PRL_10_TlBiSe2}.
The TSSs in the existing TI materials are usually deviated from a linear dispersion in going away from $E_{\rm D}$.
The ratios of $v_{g}$ at $E_{\rm F}$ with respect to that at $E_{\rm D}$ for Bi$_{2}$Te$_{2}$Se and Bi$_{2}$Se$_{2}$Te
 are found to be 1.4 and 1.6, respectively.
These values are smaller than those of the other 3D TIs (1.8 for TlBiSe$_{2}$ and 2.3 for Bi$_{2}$Se$_{3}$ though their Dirac points are deeper than in the present case), which means that the surface band dispersions for Bi$_{2}$Te$_{2}$Se and Bi$_{2}$Se$_{2}$Te
possess the wider energy range where the linearly dispersive feature is maintained above $E_{\rm D}$.

%%%%%%%%%%%%%%%%%%%%%%%%%%%%%%%%%%%%%
\begin{figure*}
\includegraphics{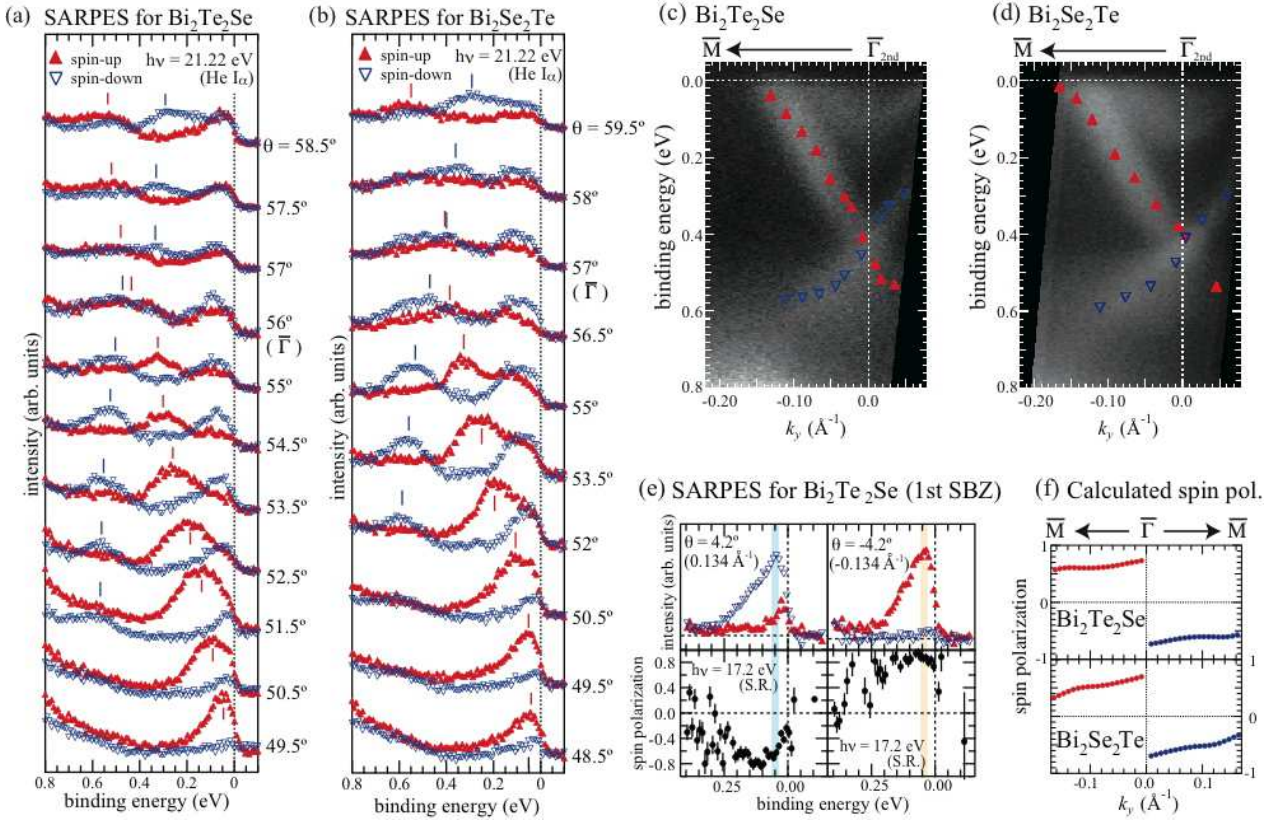}% Here is how to import EPS art
\caption{\label{fig:epsart} (a), (b) Spin-resolved energy distribution curves (EDCs) of Bi$_{2}$Te$_{2}$Se and Bi$_{2}$Se$_{2}$Te along $\overline{\Gamma \rm M}$ line obtained with unpolarized He-I$_{\alpha}$ radiation. Spin-up and spin-down intensities are denoted with triangles pointing up (red) and down (blue). (c), (d) E-k mapping ARPES measurements with  He lamp. The spin characters derived from spin-resolved spectra in Figs. 3 (a) and  3 (b) are superimposed by triangle pointing up and down. Here, $k_{y}$ is measured from $\overline{\Gamma}_{\rm 2nd}$ point. (e) Spin-resolved energy distribution curves and spin polarizations of Bi$_{2}$Te$_{2}$Se at $\theta$=$\pm$4.2$^\circ$ taken with $p$-polarized synchrotron radiation ($h\nu$=17.2~eV). (f) Theoretical spin polarization values as a function of wavenumber obtained by the first principles calculation for Bi$_{2}$Te$_{2}$Se and Bi$_{2}$Se$_{2}$Te.
 }
\end{figure*}
%%%%%%%%%%%%%%%%%%%%%%%%%%%%%%%%%%%%%

To unravel the spin character and make a quantitative analysis of spin polarization in these TSSs, we have performed the SARPES measurement.
Figures 3(a) and 3 (b) show the spin-resolved energy distribution curves (EDCs) of Bi$_{2}$Te$_{2}$Se and Bi$_{2}$Se$_{2}$Te.
Here, the spin-up and spin-down spectra are plotted with triangles pointing-up and -down, respectively.
Let us first take a look at the spin-resolved EDCs of  Bi$_{2}$Te$_{2}$Se.
A spin-up peak near $E_{\rm F}$ at $\theta$=49.5$^{\circ}$ shifts to higher $E_{\rm B}$ with increasing $\theta$.
A spin-down peak at 0.57 eV emerges from $\theta$=51.5$^{\circ}$ and move to lower $E_{\rm B}$ with increasing $\theta$.
These spin-up and -down peaks are merged at $\theta$=56$^{\circ}$  corresponding to $\overline{\Gamma} _{\rm 2nd}$ ($E_{\rm B}$=0.4~eV).
This result clearly shows that the TSS is spin split and the spin orientations are antisymmetric with respect to the $\overline{\Gamma}$ point.
For the peak near $E_F$ for $\theta$ = 53.5$^{\circ}$- 58.5$^{\circ}$, the spin-down intensity is slightly larger than that in the
spin-up channel, which might originate from final state effect \cite{Feder85}.
The observed features of band dispersion and the spin polarizations in Bi$_{2}$Se$_{2}$Te are in agreement with those in Bi$_{2}$Te$_{2}$Se as shown Fig. 3(b).
The similarity in these compounds with different stoichiometry is somewhat surprising, whose origin will be further investigated in near future.
In Fig. 3 (c) and 3 (d), we have summarized the observed spin-up and -down peaks in Figs. 3 (a) and 3 (b) with triangles pointing-up and -down, respectively.

Importantly, when there are clear peak structures in one spin channel, no apparent peak emerges in the opposite spin component over the whole energy range as shown in Figs. 3 (a) and 3 (b).
The results clearly show the presence of the surface Dirac fermions with high degree of spin polarizations in the bulk energy gap region. In fact, the TSS in Bi$_{2}$Te$_{2}$Se (Bi$_{2}$Se$_{2}$Te) maintains a high spin polarization more than 40\% (50\%) in a wide energy region across the Dirac point, though a non-polarized inelastic background contributes to the total intensity. Here, the spin polarizations are corrected from the raw data by subtracting the constant and unpolarized background from the originally derived spin-up and spin-down spectra \cite{Jozwiak11}.
In particular, the spin polarization of TSS near $E_{\rm F}$ with the smaller background contribution Bi$_{2}$Te$_{2}$Se (Bi$_{2}$Se$_{2}$Te) reaches more than 70\% (50\%).
It was, however, quite difficult to estimate the accurate values of spin polarizations near $E_{\rm F}$ in case of the measurement with He lamp due to the overlap of additional signal with considerable spin polarizations excited by the higher-energy satellite ($\beta$) line ($h\nu$=23.08 eV).
Therefore, we have tried to use the synchrotron radiation (SR) to extract accurate values.
Figure 3(e) shows the SARPES results in Bi$_{2}$Te$_{2}$Se obtained by the monochromatic SR light ($h\nu$=17.2 eV) .
As shown in the lower panel of Fig. 3 (e), the magnitude of spin polarization for the surface state reaches 87$\pm9\%$ at $\theta=-4.2^{\circ}$ and -67$\pm3\%$ at $\theta=4.2^{\circ}$.
The difference of spin polarizations in positive and negative $\theta$ is probably derived from matrix elements for optical transitions.
If the final-state spin polarizations are assumed to be equal in positive and negative $\theta$ near $\overline{\Gamma}$, an averaged spin polarization of 77$\pm5\%$ is regarded as the absolute value of the initial-state spin polarization.
Figure 3 (f) shows the theoretical spin expectation values as a function of wavenumber ($k_y$)  for Bi$_{2}$Te$_{2}$Se and Bi$_{2}$Se$_{2}$Te obtained from the first principles calculation.
 Here, we find that the theoretical spin polarization for Bi$_{2}$Te$_{2}$Se (Bi$_{2}$Se$_{2}$Te) increases in going closer to $\overline{\Gamma}$ point and it takes the maximum value of $\sim$75 \% ($\sim$70\%) at $\overline{\Gamma}$.
Note that the spin polarization can not reach 100\% because the spin
angular momentum is not a good quantum number any more due to the
strong spin-orbit entanglement as reported for Bi$_2$Te$_3$ and
Bi$_2$Se$_3$ \cite{Yazyev10}. The experimentally evaluated spin
polarization (77\%) for TSS in Bi$_2$Te$_2$Se at $k_y$ =
$\pm$0.134 \AA$^{-1}$ is larger than the theoretical value
($\sim61\%$). It is worth noting that the latter is higher in
Bi$_2$Te$_2$Se as compared with Bi$_2$Te$_3$, where reverse spin
direction was found at outer Te atom \cite{Eremeev_NatComm}. In
contrast to that in Bi$_2$Te$_2$Se the layer projected spin analysis
revealed identical spin helicity for all atomic layers. Moreover,
the spin polarization near $E_{\rm D}$ is expected to show the
higher value than those near $E_{\rm F}$ as the calculated spin
polarization implies although it is difficult to experimentally
evaluate the accurate value of spin polarization near Dirac point as
mentioned above.

In conclusion, the helical spin texture and the spin polarizations of TSS in the ternary tetradymite chalcogenides Bi$_2$Te$_2$Se and Bi$_2$Se$_2$Te have been experimentally revealed by the SARPES measurement.
The markedly high spin polarization of topological surface states has been found to be $\sim$77\% and is persistent in wide energy range across the Dirac point in those compounds.
The availability of both upper and lower TSSs promises to extend the variety of spintoronic applications, for instance, to the dual gate TI device and the topological p-n junction~\cite{Yazyev10, Wang12}.

We thank Shuichi Murakami for valuable comments.
This work was financially supported by
KAKENHI (Grant No. 20340092, 23340105), Grant-in-Aid for Scientific
Research (B) of JSPS.
K.A.K. and O.E.T. acknowledge financial support by the RSSF and RFBR (Grant No. 12-02-00226).
%\end{acknowledgments}


\end{document}